\documentstyle[12pt]{article}
\begin{document}

\author{Apalkov V.M. \\ National Research Center "Kharkov Physicothechnical
Institute", \\ 310108 Kharkov, Ukraine \\
\\
Boiko Yu.I. \\ Kharkov State University, 310077 Kharkov, Ukraine \\
\\
Carstanjen H.D. \\ Max Planck Institute of Metal Science, \\
Heisenbergstrasse 1, Stuttgard, 70569 Germany \\
\\
Slezov V.V. \\ National Research Center "Kharkov Physicothechnical
Institute", \\ 310108 Kharkov, Ukraine }
\title{Internal sinks and the smoothing of the surface structure in solids
 under irradiation.}
\maketitle

\begin{abstract}
We consider in the article the influence of the irradiation and the internal
sinks of the point defects on the rate of the flattening of the surface
structure in solids. The irradiation produces only the additional external 
sources of point defects(vacancies and interstitial atoms). The general 
system of equations is formulated. The solution of the system on the 
stationary stage of the process is found. It is shown that depending on 
the values of some parameters of the solid the irradiation can increase 
or decrease the rate of the surface flattening.
\end{abstract}

The reason of smoothing of the surface structure in solids is the difference
of the equilibrium point defect concentration on the convex and concave
parts of the surface. As a result of this difference there are the diffusion
flows of point defects which tend to redistribute the point defects
to attain the thermodinamically equilibrium state corresponding the flat
surface. The expression for the rate of surface
flattening without irradiation was received in [1]. Using this expression it
is possible to find the surface and volume diffusion coefficients analyzing
experimentally the rate of surface smoothing as a function of parameters of
the roughness of the surface [2]. In article [3] the irradiation of the
solid was introduced into the process of surface flattening. The irradiation
was considered only as a external source of point defects. It was shown in
[3] that because the depth of allocation of external source of interstitial
atoms is bigger then the depth of allocation of external source of vacancies
the irradiation increase the rate of the surface flattening. This conclusion
can be changed if we take into account the internal sinks of point defects.
Our article is devoted to this problem: the evolution of the surface
structure of solids with external sources and internal sinks of point
defects.

The main system of equations which describes the evolution of surface
structure was thoroughly analyzed in [3,4]. For small enough intensity of
irradiation we can disregard the annihilation of point defects (vacancies
and interstitial atoms). In this case we can find the contribution of
vacancies and interstitial atoms into the rate of surface flattening
separately. Let us find the contribution of vacancies.

Following the article [3] we assume that the surface of solid is described
 by the equation:
\begin{equation}
z=f(x,t)=\sum_n^{\prime }z_n(t)e^{-i\omega xn}
\end{equation}
where prime in (1) means that $n\neq 0$ in the sum (i.e. the equilibrium
surface is $z=0$), $\omega =2\pi /\lambda $, $\lambda $ is the distance
between two identical roughness on the surface, $z$-axis is directed into
the bulk of the solid. We will consider $z_n /\lambda $ as a small parameter
and will disregard the second order terms in $z_n /\lambda $ in all
equations below. Under this assumption the curvature of the surface (1)
is equal to:
\begin{equation}
K (x,t) \approx -\frac{\partial ^2f}{\partial x^2}
 = \sum_n^{\prime }z_n(t)n^2\omega
^2e^{-i\omega xn}
\end{equation}

To find the diffusion (surface and volume) flows of vacancies which modify
the surface structure in solids we should find the volume $c_{vac}$ and
the surface $u_{vac}$ vacancy concentrations as the functions of space and
time co-ordinate. We will find the concentration as the power series in
small parameter $z_n /\lambda $ to within the first-order terms in
$z_n /\lambda $. With the surface structure in the form (1) such series
are
\begin{equation}
u_{vac}(x,t)=u_{vac}^{(0)}(t)+\sum _n^{\prime }
 u_{vac}^{(n)} (t) e^{-i n \omega x}
\end{equation}
\begin{equation}
c_{vac}(x,z,t)=c_{vac}^{(0)}(z,t)+\sum _n^{\prime }
 c_{vac}^{(n)} (z,t) e^{-i n \omega x}
\end{equation}
where $c_{vac}^{(0)}$, $u_{vac}^{(0)}$ have zero order in
$z_n /\lambda $ and
 $c_{vac}^{(n)}$, $u_{vac}^{(n)}$ have the first order in  $z_n /\lambda $.

 In was shown in [3]
that under real conditions the main system of equations
describing the spatial and temporal distribution of vacancies
$c_{vac}(x,z,t)$, $u_{vac}(x,t)$ takes
the form:
\begin{equation}
\left\{
\begin{array}{l}
\frac{d u_{vac}^{(0)}}{dt} =- \nu (u_{vac}^{(0)}(t) - u_{0,vac}) +
    \frac{D_{V,vac}}{a} \left.
\frac{\partial c_{vac}^{(0)} (z,t)}{\partial z}
                        \right|_{z=f(x,t)}  \\
\frac{\partial c_{vac}}{\partial t}=D_{V,vac}\Delta
c_{vac}+I(x,z,t)-\delta _{vac}(c_{vac}-c_0)
+\frac{D_{V,vac}}{kT}\vec{\nabla } c_{vac} \vec{\nabla } U(x,z,t)\\
D_{V,vac} \left. \frac{\partial c_{vac}^{(0)} (z,t)}{\partial z}
          \right|_{z=f(x,t)}=
\left( \frac{\beta D_{V,vac}}{a} c_{vac}^{(0)}(z=f(x,t),t)-
        \frac{a}{\tau _{S}} u_{vac}^{(0)} (t) \right) \\
\left.
(c_{vac}(x,y,z,t)-c_{vac}^{(0)})\right| _{z=f(x,t)}=c_K - c_0 \\
\left. \frac{df(x,t)}{dt}\right| _{vac}=D_{S,vac}a\Delta
_S u_{vac}+a\frac{D_{S,vac}}{kT}
\vec{\nabla }_{S} (u_{vac} \vec{\nabla } _{S} U(x,z,t))+
D_{V,vac}\left. \frac{\partial c_{vac}}{\partial \vec n}\right|
_{z=f(x,y,t)}
\end{array}
\right.
\end{equation}
where index ''vac'' indicates that the quantities correspond to vacancies;
$D_{V,vac}$ is the volume diffusion coefficient of
vacancies; $D_{S,vac}$ is the surface diffusion coefficient; $\nu $
is the frequency of vacancy absorption by the surface; $I$ is the
intensity of a source of vacancies - the number of vacancies
generated by irradiation per
lattice site and per unit time; $\delta _{vac}$ is the rate of absorption of
vacancies by internal sinks ($1/\delta _{vac}$ is the vacancy mean free
time); $c_0$ is the vacancy concentration far from the surface (the
equilibrium vacancy concentration for the flat surface); $c_K=c_0+c_0\gamma
 \frac{\Omega K}{kT}$ is the equilibrium volume vacancy
concentration corresponding to the surface with the curvature $K$; $\gamma $
is the surface tension; $\Omega =a^3$, $a$ is the lattice constant; $k$ is
the Boltzman constant; $T$ is the temperature of the solid; $\Delta _S$ is
the surface part of Laplacian $\Delta =\partial _x^2+\partial _y^2+\partial
_z^2$; $\vec n$ is the inward normal to the surface of the solid;
$\beta D_{V,vac} =D^{\prime }$ determines the last jump for a vacancies
emerging from the bulk to the free surface, $\beta $ is the dimensionless
coefficient that takes into account the presence of a potential barrier
for a vacancy emerging at the sample surface ($0<\beta \leq 1$);
$\tau _S =1/ \nu $;
$\left. \frac{df(x,y,t)}{dt}\right| _{vac}$
is the contribution of vacancies into the rate of the surface smoothing;
$U(x,z,t)$ is the elastic potential energy of the vacancies in the
elastic field produced by the curved surface. Without irradiation
the terms containing $U(x,z,t)$ are very small but under some conditions
for solids under irradiation they can be important.

The potential energy $U(x,z,t)$ is given by the equation [5]:
\begin{equation}
U(x,z,t) = \left| \Omega _{vac} \right| p(x,z,t)
\end{equation}
where $\Omega _{vac}<0$ is the decreasing of the total volume of solid
under the appearing of one vacancy, $p(x,z,t)$ is the hydrodynamic
pressure in solid. For isotropic case the pressure $p(x,z,t)$ is
the solution of Laplace equation [5]:
\begin{equation}
\Delta p(x,z,t)=0
\end{equation}
with the boundary condition:
\begin{equation}
p(x,z=f(x,t)) = \gamma K(x,t)
\end{equation}
Then for the surface of type (1) the potential energy is:
\begin{equation}
U(x,z,t) = \left| \Omega _{vac}\right| \gamma
\sum _{n\neq 0} z_n (t) \omega ^2 n^2 e^{-i n \omega x}
e^{-n \omega x}
\end{equation}

We have used in the system (5) that under the condition
$\nu >> \frac{D_{S,vac}}{\lambda ^2}$ it is possible to take
$\sum_{n\neq 0} u_{vac}^{(n)} (t) exp(-i n \omega x)=u_{K}-u_{0}$ [3], where
$u_K=u_0+u_0\gamma
 \frac{\Omega K}{kT}$ is the equilibrium surface vacancy
concentration corresponding to the surface with the curvature $K$.

We shall find the quasi-stationary solution of the system (5) [3]. We have
such solution for a long experimental time $t>>\tau _{max}=max\left( \lambda
^2/D_{S,vac},l^2/D_{V,vac}\right) $ (where $l$ is the characteristic depth
at which the source of vacancies is located, $\lambda $ is distance between
two identical structure on the surface) and for a long characteristic time
of variation of roughness of the surface ($f/\dot f>>\tau _{max}$) because
in this case we can neglect the time derivative $\partial c_{vac}/\partial t$
in the first two equations of system (5) and obtain
the quasi-stationary system to
within the terms $\sim \tau _{max}/t$. The time $t$ is the parameter
at this
quasi-stationary stage. In other words, we must first determine the
quasi-steady-state fluxes for a given shape of the surface relief and then
find a closed equation determining the time variation of the surface
structure on the basis of these fluxes.

To solve the system of equations (5) we introduce the new variable:
\begin{equation}
\xi =z-f(x,t)
\end{equation}
and the new function
\begin{equation}
V(x,\xi ,t)=V^{(0)} (\xi,t) +
\sum _{n\neq 0} V^{(n)} (\xi ,t) e^{-i n \omega x}c_{vac}(x,z,t)-c_K(x,t)
\end{equation}
Then the system of equations (5) on the quasi-stationary stage takes the
form:
\begin{equation}
\left\{
\begin{array}{l}
- \nu (u_{vac}^{(0)}(t) - u_{0,vac}) + \frac{D_{V,vac}}{a} \left.
\frac{\partial V^{(0)} (\xi ,t)}{\partial \xi }
                        \right|_{\xi =0}  \\
\Delta V-f_{xx}V_\xi -2f_xV_{x\xi }=-c_{K,xx}-
\frac{I(x,\xi )}{D_{V,vac}}+\frac{\delta _{vac}}{D_{V,vac}}\left(
V+c_K-c_0\right) -\frac{1}{kT}\vec{\nabla } V \vec{\nabla } U \\
D_{V,vac} \left. \frac{\partial V^{(0)} (\xi ,t)}{\partial \xi }
          \right|_{\xi = 0}=
\left( \frac{\beta D_{V,vac}}{a} V^{(0)}(\xi =0,t)-
        \frac{a}{\tau _{S}} (u_{vac}^{(0)} (t)-u_{0}) \right) \\
 V^{(n)}(x,\xi =0)=0 \\
\left. \frac{df(x,t)}{dt}\right| _{vac}=D_{S,vac}a\Delta
_S u_{vac}+a\frac{D_{S,vac}}{kT}
\vec{\nabla }_{S} (u_{vac} \vec{\nabla } _{S} U(x,\xi ,t))+
D_{V,vac}\left. \frac{\partial V}{\partial \xi }\right|
_{\xi = 0}
\end{array}
\right.
\end{equation}

We take the source of vacancies in the form of $\delta $-function:
\begin{equation}
I(x,\xi ) = I_0 \Omega \delta (\xi - l)
\end{equation}
where $I_0$ is the number of vacancies emerging at the surface per unit time
and per unit area. When the sample is bombarded by ions, the profile of
generated point defects has a Gaussian form, and (7) is a good approximation
for a narrow distribution of defects.

We find the solution of the first four equation of the system (12)
with the accuracy to within
$z_n/\lambda $ [3]. Then we substitute this solution into the fifth equation
of the system (12) and find the contribution of vacancies into the rate of
the surface smoothing:
\begin{equation}
\begin{array}{lrl}
\left. \frac{dz_n(t)}{dt}\right| _{vac} & = &
 -z_n(t)D_{S,vac}a
\frac{c_0\gamma \Omega }{kT}\omega ^4n^4-z_n(t)D_{V,vac}\frac{c_0\gamma
\Omega }{kT}\frac{w^2n^2}{\mu _{vac}}\left( \omega ^2n^2+q_{s,vac}^2\right)
- \\
 & &  -D_{S,vac}
\frac{I_o \Omega }{\nu }\frac{\gamma \left| \Omega_{vac} \right|}{kT}
z_n(t) \omega^4 n^4 e^{-l q_{s,vac}}   \\
 &  & +I_0\Omega \mu _{vac}z_n(t)\left( e^{-lq_{s,vac}}-e^{-\mu
_{vac}l}\right)
\end{array}
\end{equation}
where $q_{s,vac}=\sqrt{\frac{\delta _{vac}}{D_{V,vac}}}$, $\mu _{vac}=%
\sqrt{\omega ^2n^2+q_{s,vac}^2}$ and the condition $aq_s /\beta $ was used.

For interstitial atoms the system of equation (12) has the form [3]:
\begin{equation}
\left\{
\begin{array}{l}
- \nu u_{in}^{(0)}(t)  + \frac{D_{V,in}}{a} \left.
\frac{\partial V^{(0)} (\xi ,t)}{\partial \xi }
                        \right|_{\xi =0} = 0  \\
\Delta V-f_{xx}V_\xi -2f_x V_{x\xi }= -
\frac{I(x,\xi )}{D_{V,in}}+\frac{\delta _{in}}{D_{V,in}}
V -\frac{1}{kT}\vec{\nabla } V \vec{\nabla } U \\
D_{V,in} \left. \frac{\partial V^{(0)} (\xi ,t)}{\partial \xi }
          \right|_{\xi = 0}=
\left( \frac{\beta D_{V,in}}{a} V^{(0)}(\xi =0,t)-
        \frac{a}{\tau _{S}} u_{in}^{(0)} (t) \right) \\
 V^{(n)}(x,\xi =0)=0 \\
\left. \frac{df(x,t)}{dt}\right| _{in}=a\frac{D_{S,in}}{kT}
\vec{\nabla }_{S} (u_{in} \vec{\nabla } _{S} U(x,\xi ,t))+
D_{V,in}\left. \frac{\partial V}{\partial \xi }\right|
_{\xi = 0}
\end{array}
\right.
\end{equation}
where we have taken into account that $c_{0,in}\approx 0$ (i.e.
$c_{0,in}<<c_{0,vac}$); the index ''$in$'' in system (15) indicates that the
quantities correspond to interstitial atoms. Solving the system (15) we find
the contribution the interstitial atoms into the rate of the surface
flattening:
\begin{equation}
\left. \frac{dz_n(t)}{dt}\right| _{in}=-I_0\Omega \mu _{in}z_n(t)\left(
e^{-l_{in}q_{s,in}}-e^{-\mu _{in}l}\right)
-D_{S,in}
\frac{I_o \Omega }{\nu }\frac{\gamma \left| \Omega_{in} \right|}{kT}
z_n(t) \omega^4 n^4 e^{-l_{in} q_{s,in}}
\end{equation}
where $q_{s,in}=\sqrt{\frac{\delta _{in}}{D_{V,vac}}}$, $\mu _{in}=\sqrt{%
\omega ^2n^2+q_{s,in}^2}$.

Then the total rate of surface smoothing is:
$$
\begin{array}{lcl}
\frac{dz_n(t)}{dt} & = & -z_n(t)D_{S,vac}a
\frac{c_0\gamma \Omega }{kT}\omega ^4n^4-z_n(t)D_{V,vac}\frac{c_0\gamma
\Omega }{kT}\frac{w^2n^2}{\mu _{in}}\left( \omega ^2n^2+q_{s,vac}^2\right) -
\\
 &  & -D_{S,vac}
\frac{I_o \Omega }{\nu }\frac{\gamma \left| \Omega_{vac} \right|}{kT}
z_n(t) \omega^4 n^4 e^{-l_{vac} q_{s,vac}}
-D_{S,in}
\frac{I_o \Omega }{\nu }\frac{\gamma \left| \Omega_{in} \right|}{kT}
z_n(t) \omega^4 n^4 e^{-l_{in} q_{s,in}}  \\
 &  & -I_0z_n(t)\Omega \left\{ \mu _{in}\left(
e^{-l_{in}q_{s,in}}-e^{-\mu _{in}l_{in}}\right) -\mu _{vac}\left(
e^{-l_{vac}q_{s,vac}}-e^{-\mu _{vac}l_{vac}}\right) \right\}
\end{array}
$$
and after integration we find:
\begin{equation}
\begin{array}{lcl}
\ln \frac{z_n(t)}{z_n(0)} & = & -D_{S,vac}a
\frac{c_0\gamma \Omega }{kT}\omega ^4n^4t-D_{V,vac}
\frac{c_0\gamma \Omega }{kT}\frac{w^2n^2}{\mu _{in}}\left(
 \omega ^2n^2+q_{s,vac}^2\right) t- \\
 &  & -D_{S,vac}
\frac{I_o \Omega }{\nu }\frac{\gamma \left| \Omega_{vac} \right|}{kT}
 \omega^4 n^4 e^{-l_{vac} q_{s,vac}}t
-D_{S,in}
\frac{I_o \Omega }{\nu }\frac{\gamma \left| \Omega_{in} \right|}{kT}
 \omega^4 n^4 e^{-l_{in} q_{s,in}}t  \\
 & & -I_0\Omega \left\{ \mu _{in}\left( e^{-l_{in}q_{s,in}}-e^{-\mu
_{in}l_{in}}\right) -\mu _{vac}\left( e^{-l_{vac}q_{s,vac}}-e^{-\mu
_{vac}l_{vac}}\right) \right\} t
\end{array}
\end{equation}

The first and the second terms in equation (17) describe the evolution of
the surface without radiation. For $q_{s,vac}=0=q_{s,in}$ (no internal
sinks) these terms were received in [1]. The last three terms in (17)
is the contribution of irradiation into $\ln \frac{z_n(t)}{z_n(0)}$. For
$q_{s,vac}=0=q_{s,in}$ we rederive the result of article [3].

For $q_{s,vac}<<\omega n$ ($q_{s,in}<<\omega n$), $q_{s,vac}<<1/l_{vac}$ ($%
q_{s,in}<<1/l_{in}$) and $\omega l_{vac}<<1$ ($\omega l_{in}<<1$) we find
from the equation (17) the corrections to the results of [1] and [3]:
$$
\ln \frac{z_n(t)}{z_n(0)}=-D_{S,vac}a\frac{c_0\gamma \Omega }{kT}\omega
^4n^4t-D_{V,vac}\frac{c_0\gamma \Omega }{kT}\omega ^3n^3\left( 1+
\frac{q_{s,vac}^2}{2\omega ^2n^2}\right) t-
$$
$$ -D_{S,vac}
\frac{I_o \Omega }{\nu }\frac{\gamma \left| \Omega_{vac} \right|}{kT}
 \omega^4 n^4 (1-l_{vac} q_{s,vac}) t
-D_{S,in}
\frac{I_o \Omega }{\nu }\frac{\gamma \left| \Omega_{in} \right|}{kT}
 \omega^4 n^4 (1-l_{in} q_{s,in}) t -
$$
\begin{equation}
-I_0\Omega \omega ^2n^2\left\{ l_{in}\left( 1-\frac{q_{s,in}}{\omega n}+%
\frac{q_{s,in}^2}{2\omega ^2n^2}\right) -l_{vac}\left( 1-\frac{q_{s,vac}}{%
\omega n}+\frac{q_{s,vac}^2}{2\omega ^2n^2}\right) \right\} t
\end{equation}

We see that we have the second order correction $q_{s,vac}^2/\omega ^2$ to
the result without sinks and without irradiation (the second term in
equation (18) [1]) and the first order correction $q_{s,vac}/\omega $,
 $l_{vac} q_{s,vac}$
($q_{s,in}/\omega $, $l_{in} q_{s,in}$ ) to the result
with irradiation and without sinks [3].

The sign of the last term in eq. (17) depends on the sign of the
expression:
\begin{equation}
\psi = l_{in} \left( 1-\frac{q_{s,in}}{\omega n}+\frac{q_{s,in}^2}{2\omega
^2n^2} \right) - l_{vac} \left( 1-\frac{q_{s,vac}}{\omega n}+\frac{%
q_{s,vac}^2}{2\omega ^2n^2} \right)
\end{equation}
For $\psi > 0 $ the last term is negative ant it
increases the rate of the surface smoothing
and for $\psi < 0 $ the last term increases the roughness of the surface.
The leading term in the third term have $\omega ^2$-dependence on $\omega $.

For $q_{s,vac}>>\omega $ ($q_{s,in}>>\omega $) the equation (17) takes the
form:
$$
\ln \frac{z_n(t)}{z_n(0)}=-D_{S,vac}a\frac{c_0\gamma \Omega }{kT}\omega
^4n^4t-D_{V,vac}\frac{c_0\gamma \Omega }{kT}q_{s,vac}\omega ^2n^2t-
$$
$$
 -D_{S,vac}
\frac{I_o \Omega }{\nu }\frac{\gamma \left| \Omega_{vac} \right|}{kT}
z_n(t) \omega^4 n^4 e^{-l_{vac} q_{s,vac}}
-D_{S,in}
\frac{I_o \Omega }{\nu }\frac{\gamma \left| \Omega_{in} \right|}{kT}
z_n(t) \omega^4 n^4 e^{-l_{in} q_{s,in}} -
$$
\begin{equation}
-I_0\Omega \frac{\omega ^2n^2}2\left\{
l_{in}e^{-l_{in}q_{s,in}}-l_{vac}e^{-l_{vac}q_{s,vac}}\right\} t%
\end{equation}
where the conditions $l_{vac}\omega ^2<<q_{s,vac}$ and $l_{vac}\omega
^2<<q_{s,vac}$ were assumed. In this case the second and the last terms in
equation (19) have the same $\omega ^2$ dependence on the frequency
$\omega $. The sign of the last term is determined by the expression:
\begin{equation}
\psi _1=l_{in}e^{-l_{in}q_{s,in}}-l_{vac}e^{-l_{vac}q_{s,vac}}
\end{equation}
Depending on the values of $q_{s,in}$ ($q_{s,vac}$) and $l_{in}$ ($l_{vac}$)
the irradiation tends to decrease (for $\psi _1>0$) or to increase (for $%
\psi _1<0$) the roughness of the surface. The expression (15) shows that the
interstitial atoms tends to make the surface become flat and the vacancies
created by irradiation tends to increase the roughness of the surface. The
influence of the point defects on the evolution of the surface is determined
by the value of $le^{-lq_s}$.For example, if the interstitial atoms are
absorbed by internal sinks stronger that the vacancies are then the
vacancies are the main defects and the expression $\psi _1$ is negative -
the irradiation increases the roughness of the surface.

In conclusion, we derived the expression for the rate of the surface
smoothing taking into account the external sources of point defects (due to
irradiation) and the internal sinks of point defects.
We have found that the irradiation has tow contribution into the rate of
surface smoothing: the first contribution (the third and the fourth
terms in equation (17)) increase the rate of the surface flattening and
the influence of the second contribution of irradiation
(the last term in equation (17))
on the evolution of the surface structure depends on the
sign of the expression
$\psi =l_{in}\left( 1-\frac{q_{s,in}}{\omega n}+
\frac{q_{s,in}^2}{2\omega ^2n^2}\right) -l_{vac}\left( 1-\frac{
q_{s,vac}}{\omega n}+\frac{q_{s,vac}^2}{2\omega ^2n^2}\right) $
(for $q_{s,vac}<<\omega $, $q_{s,in}<<\omega $ and $q_{s,vac}l_{vac}<<1$,
$q_{s,vac}l_{vac}<<1$) and the expression $\psi
_1=l_{in}e^{-l_{in}q_{s,in}}-l_{vac}e^{-l_{vac}q_{s,vac}}$
(for $q_{s,vac}>>\omega $, $q_{s,in}>>\omega $).
For $\psi >0$ ($\psi _1>0$) the second contribution of
irradiation increases the rate of the surface flattening and for $\psi <0$
($\psi _1<0$) the it increases the roughness of the surface.

\begin{center}
{\bf Literature}
\end{center}

   [1] W.W. Mullins. J. appl. Phys. {\bf 30}, p. 77 (1959).

   [2] K. Hoehne, R. Sizmann. Phys. Stat. Sol. {\bf 5}, p. 577 (1971).

   [3] V.V. Slezov, Yu.I. Boiko, V.M. Apalkov, H.D. Carstanjen.
 Journal of Low Temperature Physics, {\bf 23}, p. 97 (1997).

   [4] V.V. Slezov, Yu.I. Boiko, V.M. Apalkov, H.D. Carstanjen, in press.

   [5] L.D. Landau, E.M. Lifshit, Teoriya Uprugosty (Theory of Elasticity),
Moscow, Nauka (in Russian) (1965).

\end{document}